\documentclass{article}
\pdfoutput=1

\usepackage{spconf,amsmath,graphicx,amsfonts}
\usepackage{cite}
\usepackage{multirow}
\usepackage{caption}
\usepackage{subcaption}
\usepackage{hyperref}
\usepackage[T1]{fontenc}

\def\R{\mathbb{R}}
\newcommand{\Nms}[2]{\mathcal{N}(#1,#2 \mathbf{I})}
\def\N{\Nms{\mathbf{0}}{}}

\def\y{\mathbf{y}}
\def\x{\mathbf{x}}
\def\z{\mathbf{z}}
\def\e{\mathbf{\epsilon}}

\def\s{\sigma}
\def\ve{\varepsilon_\eta}

\def\xx{\underline{\x}}
\def\zz{\underline{\z}}
\def\ee{\underline{\e}}

\def\a{\underline{\mathbf{a}}}
\def\T{^\mathsf{T}}

\def\X{\mathbf{X}}
\def\Xt{\tilde{\X}}

\def\xt{\tilde{\x}}
\def\xxt{\tilde{\xx}}

\newcommand{\pr}[1]{p(#1)}
\newcommand{\pc}[2]{\pr{#1 | #2}}

\def\grad{\nabla}

\title{DISTRIBUTION PRESERVING SOURCE SEPARATION WITH TIME FREQUENCY PREDICTIVE MODELS}
\name{Pedro~J.~Villasana~T. \qquad Janusz~Klejsa \qquad Lars~Villemoes \qquad Per~Hedelin}
\address{Dolby Sweden AB, Stockholm, Sweden}

\begin{document}

\maketitle

\begin{abstract}
We provide an example of a distribution preserving source separation method, which aims at addressing perceptual shortcomings of state-of-the-art methods. Our approach uses unconditioned generative models of signal sources. Reconstruction is achieved by means of mix-consistent sampling from a distribution conditioned on a realization of a mix. The separated signals follow their respective source distributions, which provides an advantage when separation results are evaluated in a listening test.
\end{abstract}


\section{Introduction}
\label{sec:intro}

Reconstructions of source signals generated by state-of-the-art audio separation algorithms often sound unnatural. Even a scheme with good objective performance may fall short in a listening test, since its reconstruction may be deemed non-plausible by a human observer. This problem is also discussed in the context of signal enhancement, where a trade-off between distortion and plausibility of the reconstruction is recognized \cite{blau2018perception}. In this paper, we provide an example of a distribution preserving source separation (DPSS) scheme, where we require that the reconstructed signal components obey their respective source distributions.
\par Once a mixture of audio signals is created, some information about its components is inevitably lost. Regression-based separation schemes \cite{guso2022loss} tend to suppress reconstruction when there is an uncertainty about the source. This can, for example, introduce spectral holes, which can be sometimes easily detected while listening, even though the objective performance is high \cite{luo2018tasnet}.
\par Generative separation methods were introduced in \cite{subakan2018generative, narayanaswamy2020unsupervised, hawthorne2022multi}, where schemes operate on magnitude spectra. Such a setting is generally suboptimal, since it discards the phase information, which may introduce artifacts \cite{lluis2018end}. A post-processing approach to improve the perceptual performance of spectrogram-based separation was proposed in \cite{schaffer2022music}. In the context of audio separation, generative modeling in the signal domain currently provides the state-of-the-art reconstruction quality \cite{jayaram2021parallel, mariani2023multi}. \par
\begin{figure*}[h!]
\begin{minipage}{.33\textwidth} 
  \includegraphics[width=.99\textwidth]{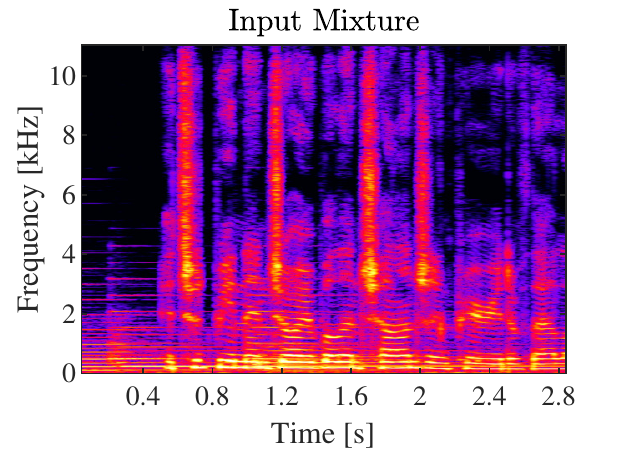}\vspace{-0.3cm}\hfill
\end{minipage}  
\begin{minipage}{.33\textwidth}  
  \includegraphics[width=.99\textwidth]{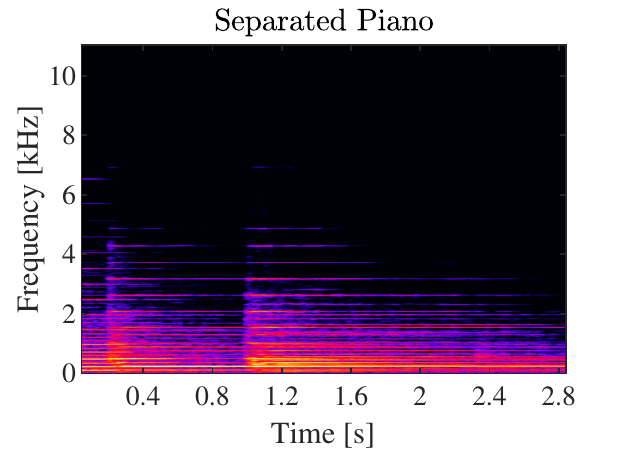}\vspace{-0.3cm}\hfill
\end{minipage}  
\begin{minipage}{.33\textwidth}   
  \includegraphics[width=.99\textwidth]{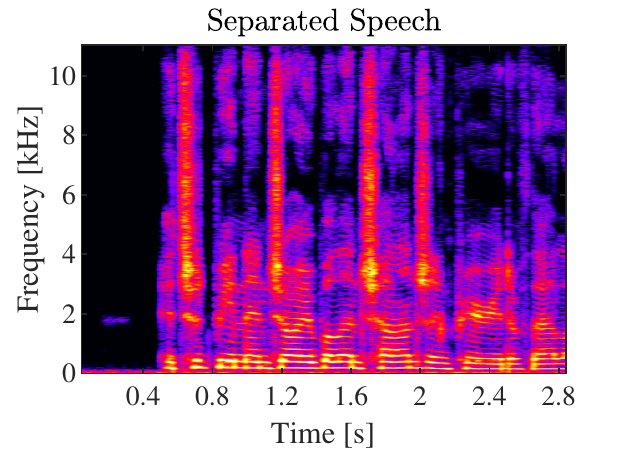}\vspace{-0.3cm}
\end{minipage}
  \caption{Spectrograms: (left) Input mixture (speech+piano); (center) Separated piano (DPSS);  (right) Separated speech (DPSS). \label{fig:spectrograms}}
\end{figure*}

 The approach proposed in this paper stems from \cite{jayaram2021parallel}, which uses generative unconditional models of signal components in the construction of a separation scheme. While the separation scenarios of \cite{jayaram2021parallel, mariani2023multi} are toy-like (because these approaches assume that generative models of all the mixture components are available), they still provide a clean testing ground for the proposed separation approach. In our case, we focus on employing high-quality generative models to provide in-distribution reconstructions. Specifically, we use generative models operating in a quadrature modulated filter (QMF) bank domain \cite{Vaidyanathan1994} that were designed to deliver high quality audio synthesis. Our goal is to make the separated components perceptually indistinguishable from observations of signals generated from their sources (as illustrated in Fig.~\ref{fig:spectrograms}). We choose separation of piano-speech mixtures to facilitate comparison with \cite{jayaram2021parallel}, but also to demonstrate that the principle works with diverse signal categories.
\par There are many objective performance measures for evaluating source separation methods \cite{cano2016evaluation, kastner2019efficient, le2019sdr}. However, it is also commonly acknowledged that such measures do not reliably predict the perceptual performance \cite{torcoli2021objective}. Here we focus on evaluating the separation results in a listening test. We use the MUSHRA methodology \cite{bs.1534-3}, as normally applied for evaluation of audio coding schemes. 
\par The paper is organized as follows. We describe the separation method in Section \ref{sec:separation}. The approach involves usage of source models capable of providing in-distribution reconstructions of the sources. We describe a structure of such a model in Section \ref{sec:model}. The results of objective evaluation of the scheme and results of the listening tests are provided in Section \ref{sec:eval}.
\section{Separation method}
\label{sec:separation}
An observation of a downmix $\y = g(\xx)$ where $g: \R^{S \times N} \rightarrow \R^{N}$ is the result of a mixing operation of $S$ independent components contained in a realization of a source vector $\xx$. The goal is to reconstruct the components in $\xx$ by sampling from a trained model $\pc{\xx}{\y}$.  We follow the approach of \cite{jayaram2021parallel}, where Langevin sampling facilitates working with the log gradient of the posterior of the joint distribution: 
\begin{equation}
\grad_{\xx}\log p(\xx,{\y})=\grad_{\xx} \log p({\y|\xx}) + \grad_{\xx} \log p(\xx).
\label{eq:scores}
\end{equation}
Such a sampling scheme can be implemented by using a score matching model \cite{song2020score, pascual2022full, mariani2023multi}, where $\grad_{\xx} \log p(\xx)$ would be the score models for the components of the mix (which would be the only trainable parts of the scheme). Alternatively, the gradients can be computed in runtime by using trained generative models $p(\xx)$. 

\subsection{Langevin sampling}
Due to \eqref{eq:scores}, the Langevin sampling for sufficiently small $\eta$ can be implemented as:
\begin{align} \label{eq:lgv}
&\xx_{t_{i}} = \xx_{t_{i-1}} + \eta \grad_{\xx} \log \pc{\xx_{t_{i-1}}}{\y} + \sqrt{2\eta} \ee_{t_{i}} \\
&= \xx_{t_{i-1}} + \eta \grad_{\xx} \left( \log \pr{\xx_{t_{i-1}}} + \log \pc{\y}{\xx_{t_{i-1}}} \right) + \sqrt{2\eta} \ee_{t_{i}} \nonumber
\end{align}
where $i = 1, ..., I$ is the iteration number, $t \in \mathcal{U}(0,1)$ describes the algorithmic time going from $t_{0} = 0$ to $t_{I} = 1$, and $\ee \in \R^{S \times N}$ is a vector of S noise realizations sampled from $\N$. The scheme of \cite{jayaram2021parallel} used auto-regressive models to describe $\pr{\xx}$. However, in that case $\xx$ was discrete and therefore non-differentiable. This problem was alleviated by sampling from a smoothed version of $\xx$, where $\xxt = \xx + \s_m \zz$ and $\zz \in \R^{S \times N}$ consisting of S noise realizations sampled from $\N$, for $\s_{1} > \dots > \s_{M}$. This requires using a collection of $M$ models for each source. Here, instead, we use noisy-predictors described in Section \ref{sec:noisy_predictors}, where there is only a single model per source.

\subsection{Auto-regressive source model}
While the separation approach is compatible with modern score matching models such as \cite{pascual2022full},\cite{mariani2023multi}, here, we use auto-regressive (AR) modeling of signals (as in \cite{jayaram2021parallel}), since they seem to have good enough performance for our task. However, our AR models operate in a filter-bank domain on a continuous signal representation. It is known that time-domain AR modeling of signals close to being a sum of AR processes (e.g., piano) is not efficient \cite{granger1976time}. The decorrelating property of the filter-bank seems to aid the model efficiency for such complex sources in our experiments.

In our model, a probability distribution of a random variable represented by a matrix of time-frequency tiles of a signal representation provided by the filter-bank is given by a product of probabilities of conditionally independent frames:
\begin{align}
\pr{\X} = \prod_{n} \pc{\x_{n}}{\X_{<n}},
\end{align}
where $\x_{n}$ denotes the frequency coefficients at time frame $n$ and $\X_{<n}$ comprises past time frames. This model structure resembles an MDCT model of \cite{davidson2022high}, but in this work we use another implementation designed for a real-valued QMF. We discuss implementation of the model in Section \ref{sec:model}.

\subsection{Noisy predictor}
\label{sec:noisy_predictors}
The Langevin sampling method depends on providing a description of $\pr{\x_{t_{i}}}$. This means that we must have models that are able to describe the probability of $\x$ in between $\x_{0}$ (noise) and $\x_{1}$ (source). Here we perform AR modelling for noisy sources $\xt = \x + \s \z$ with $\z \sim \N$ conditioned on the noise level, i.e.~$\pc{\xt}{\s}$. This helps to have more control over the Langevin sampling algorithm, thus modifying \eqref{eq:lgv} to
\begin{align} \label{eq:lgv_cond}
&\xxt_{t_{i}} = \xxt_{t_{i-1}} + \eta \grad_{\xxt} \log \pc{\xxt_{t_{i-1}}}{\y,\s_{t_{i}}} + \sqrt{2\eta} \ee_{t_{i}} \\
&\quad = \xxt_{t_{i-1}} + \eta \grad_{\xxt} \left( \log \pc{\xxt_{t_{i-1}}}{\s_{t_{i}}} + \log \pc{\y}{\xxt_{t_{i}},\s_{t_{i}}} \right) + \nonumber \\
&\qquad+ \sqrt{2\eta} \ee_{t_{i}}. \nonumber
\end{align}
Writing the linear function $g$ as $g(\xx) = \a\T \xx$ we have that
\begin{align}
\y = g(\xx) = g(\xxt - \s \zz) \sim \Nms{g(\xxt)}{\s^2 \|\a\|^2},
\end{align}
where $\a \in \R^{S}$ defines the weights for mixing the sources. The proposed construction eliminates the need for training a sequence of de-noising models as in \cite{jayaram2021parallel}.

\begin{figure*}[ht!]
\vspace{-0.3cm}
\centering
\begin{subfigure}[b]{.33\textwidth}
  \centering
  \includegraphics[scale=0.5]{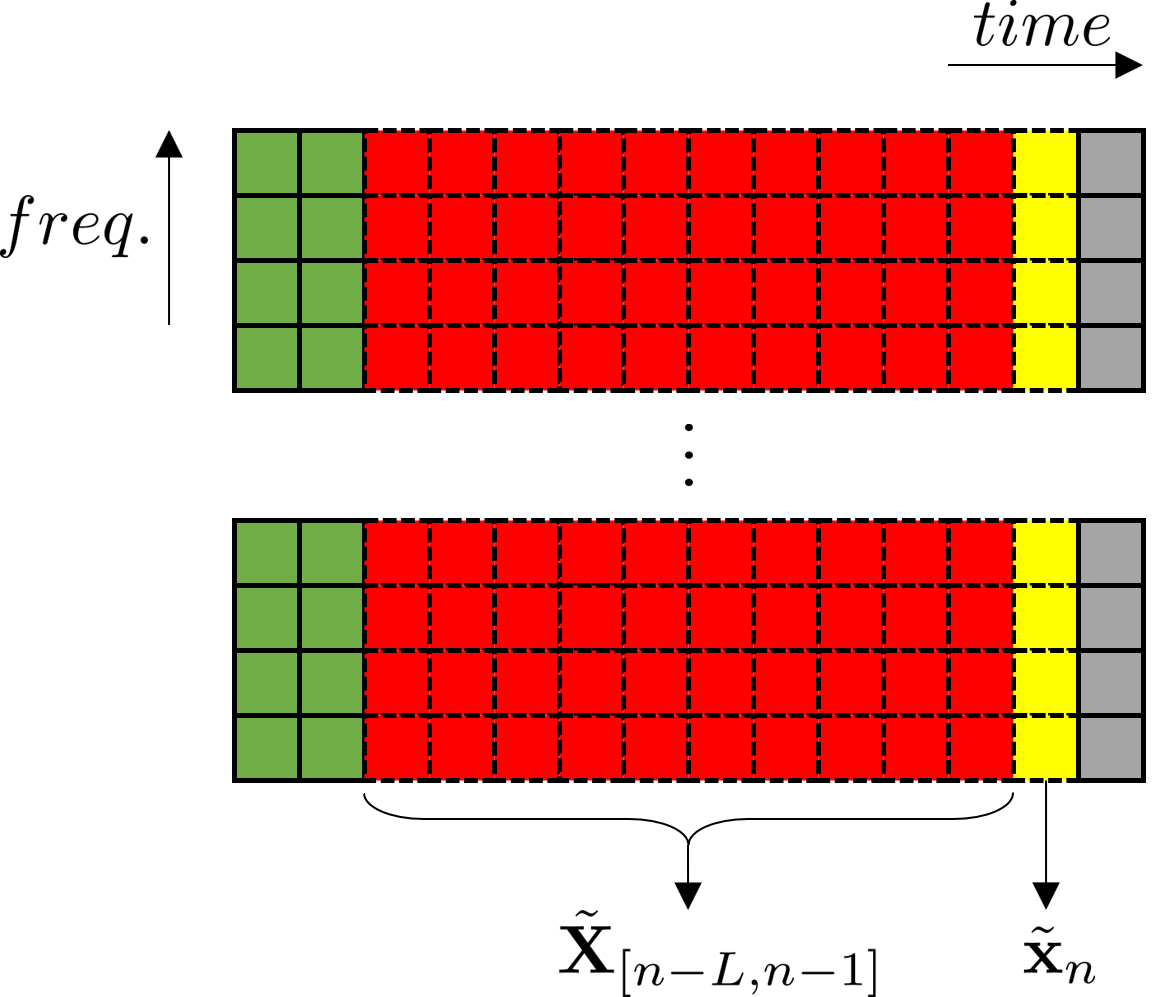}
  \vspace{0.5cm}
  \caption{Generation example}
  \label{fig:model_diag}
\end{subfigure}
\begin{subfigure}[b]{.33\textwidth}
  \centering
  \includegraphics[scale=0.5]{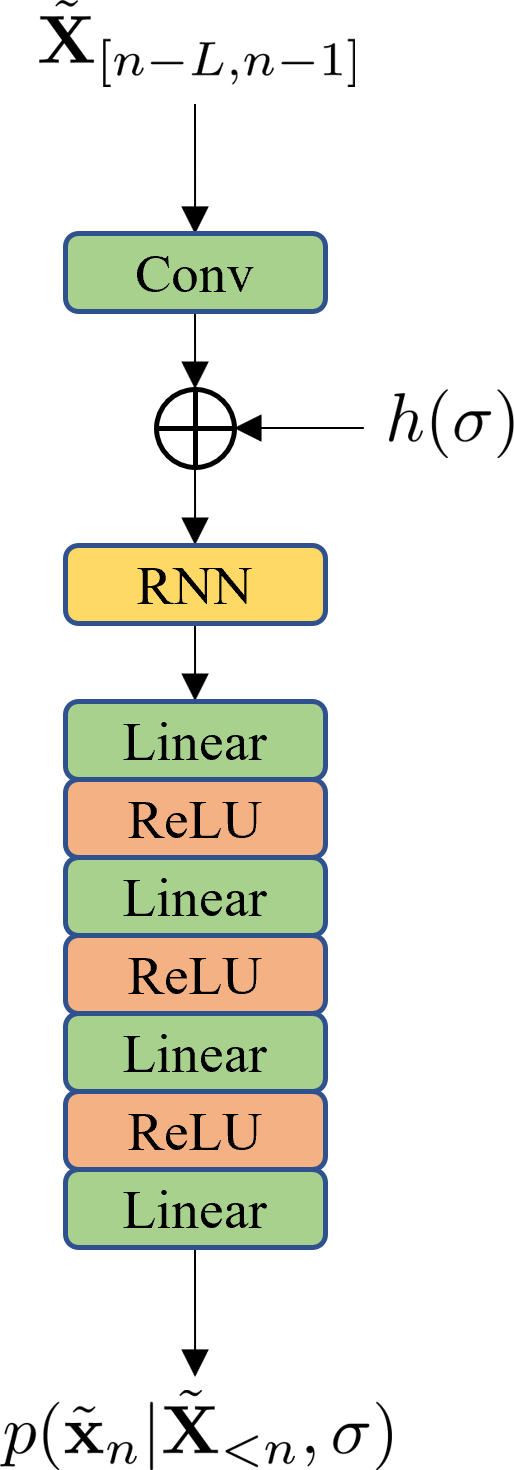}
  \caption{Prediction network}
  \label{fig:model_ar}
\end{subfigure}
\begin{subfigure}[b]{.33\textwidth}
  \centering
  \includegraphics[scale=0.5]{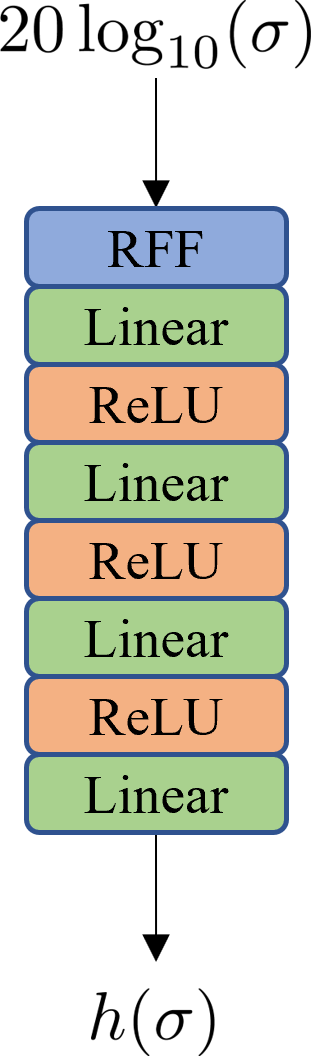}
  \vspace{0.5cm}
  \caption{Conditioning network}
  \label{fig:model_cond}
\end{subfigure}
  \caption{Source model operation.}
  \label{fig:model}
\end{figure*}

\subsection{Scheduler}
Usually the steps $\s_{t_0} > \s_{t_1} > \cdots > \s_{t_I}$ in \eqref{eq:lgv_cond} are geometrically spaced by $\gamma = \s_{t_{i+1}} / \s_{t_i}$ with $t_0 = 0$ and $t_I = 1$. The ALS scheduler \cite{song2019generative} is equivalent to apply \eqref{eq:lgv_cond} with $\eta = \ve \s_{t_i}/\s_{t_I} = \ve \gamma^{i-I}$ where $ \ve$ is a hyperparameter.

The re-parameterized CAS scheduler introduced in \cite{serra2021tuning} applies the Langevin procedure as
\begin{align} \label{eq:lgv_cas}
&\xxt_{t_{i}} = \xxt_{t_{i-1}} + \alpha \s^2_{t_{i}} \grad_{\xxt} \log \pc{\xxt_{t_{i-1}}}{\y,\s_{t_{i}}} + \beta \s_{t_{i+1}} \ee_{t_{i}} \\
&\quad = \xxt_{t_{i-1}} + \beta \s_{t_{i+1}} \ee_{t_{i}} +  \nonumber \\
&\qquad+ \alpha \s^2_{t_{i}} \grad_{\xxt} \left( \log \pc{\xxt_{t_{i-1}}}{\s_{t_{i}}} + \log \pc{\y}{\xxt_{t_{i-1}},\s_{t_{i}}} \right), \nonumber
\end{align}
where $\alpha = 1 - \gamma^\eta$, $\beta = \sqrt{1 - \gamma^{2(\eta-1)}}$ with $\eta \geq 1$ for $i \in \{1, 2, \dots, I-1\}$, and $\alpha = 1$, $\beta = 0$ for $i = I$ ($t = 1$).

\section{Source Model}\label{sec:model}
We designed an auto-regressive model that operates in the QMF domain, predicting noisy frames conditioned on the noise level, i.e.~the model performs
\begin{align}
\pc{\Xt}{\s} = \prod_{n} \pc{\xt_{n}}{\Xt_{<n}, \s}
\end{align}
with $\Xt$ being the result of adding Gaussian noise of power $\s^2$ to the clean signal $\X$. This is a simplified version of the model in \cite{davidson2022high} where they do frequency prediction within a time frame. Here, we predict the whole frame in one shot.

A high-level view on the model architecture is shown in Fig.~\ref{fig:model}. Here, we highlight the two aspects of the model: its predictive structure and noise-level conditioning.

\textbf{Prediction}. Most natural occurring audio signals have certain correlation in time and frequency, examples of these are the tempo of a song and the timbre of a musical instrument, respectively. To exploit this correlation we designed the model to perform prediction of vectors. Looking at Fig.~\ref{fig:model_diag}, to predict the samples $\xt_{n}$ (yellow) in frame $n$, the model takes $L$ frames in the past $\Xt_{[n-L,n-1]}$ (red). These samples are passed through a convolutional layer and the output is added to the output of the conditioning network. The resulting sum goes into an RNN that handles coherence over time. Then the output of the RNN is put through an MLP layer consisting of 4 linear layers with ReLUs in between them. The output of the final linear layer contains the parameters necessary to create the multivariate distribution $\pc{\xt_{n}}{\Xt_{<n}, \s}$. The prediction network is shown in Fig.~\ref{fig:model_ar}.

\textbf{Noise-level conditioning}. The conditioning network in Fig.~\ref{fig:model_cond} takes the noise level $\s$ in dB and converts it into Random Fourier Features (RFF) \cite{rouard2021crash}. These features are then passed through a 4-layer MLP with ReLUs in between each linear layer.
\vspace{-0.5cm}
\section{Evaluation}
\label{sec:eval}
We evaluated the proposed method on mixtures created from piano and speech signals. The piano signals came from the Supra database \cite{shi2019supra}, while the speech excerpts came from the VCTK database \cite{yamagishi2019cstr}. This selection allows us to demonstrate the gain in objective and subjective performance, when the separation framework of \cite{jayaram2021parallel} is configured towards source distribution preservation by employing high-quality source models.
\subsection{Model setup and training}
We used a QMF bank of 64 channels. The model was designed to predict 1 frame at a time by looking at 10 frames in the past ($L=10$ in Fig.~\ref{fig:model_diag}). All elements in the model use a hidden dimension of size 1024 and the output stage of the last linear layer in Fig.~\ref{fig:model_ar} produced 128 parameters. These parameters represent means $\mu$ and scales $s$ used to construct Logistic distributions $\text{Log}(x; \mu, s) = \frac{1}{4s} \text{sech}^2 \left( \frac{x-\mu}{2s} \right)$ for each of the 64 QMF channels in the frame. We opted for using the Logistic distribution due to its smooth and range-limited derivative $\frac{\partial}{\partial x} \log \text{Log}(x; \mu, s) = -\frac{1}{s} \text{tanh} \left( \frac{x-\mu}{2s} \right)$. This property proved beneficial when computing $\grad_{\Xt} \log \pc{\Xt}{\s}$ for the application of \eqref{eq:lgv_cas}.

We trained source models using the VCTK speech and Supra piano datasets. Each model had approximately 17M parameters. The datasets were randomly divided into training, validation and test sets, consisting of 80-10-10 percent of the data, respectively. All items were adjusted to a length of 8~seconds by concatenation or truncation. We trained each model using NLL loss for 1M iterations, using the ADAM optimizer \cite{kingma2014adam}, starting with a learning rate of 1e-4 and reducing it with a cosine scheduler \cite{loshchilov2016sgdr} to a final value of 1e-6. Each iteration consists of a batch of 64 items, using a sequence length of 1~second. Every file in the batch was corrupted by adding Gaussian noise with power distributed uniformly in the range [-90, 0] dB. We kept this noise level for the entire duration of the file, such that 8 consecutive iterations belonged to the same items and noise levels.

When performing source separation, we used the CAS scheduler \eqref{eq:lgv_cas} starting from $\s_{0} = 0$ dB to $\s_{1} = -90$ dB, using $\eta = 90$ for $I = 1500$ iterations. We noted that the results varied significantly depending on the power level of the individual sources. Therefore, we decided to normalize the mix to a level of -23~dB power before applying \eqref{eq:lgv_cas}, and de-normalize the estimated sources afterwards.
\vspace{-0.35cm}
\subsection{Objective performance}
We evaluated the separation performance by using SI-SDR \cite{le2019sdr} as an objective measure. We constructed a regression-based benchmark with an ideal ratio mask (IRM) computed on per-bin basis with a 1024 samples stride. We used a time-domain scheme of \cite{jayaram2021parallel} (PNF), which is based on WaveNet \cite{OordDZSVGKSK16} operating with a discrete output stage (8-bit PCM), where we computed the SI-SDR with respect to the original signal (16-bit PCM). Finally, we included the proposed scheme. Results of these three approaches are shown in Table \ref{tab1} for speech, piano, and the mix. It can be seen that the proposed scheme has a superior performance for both source categories. But as expected, an ideal regression scheme has the best mix consistency. The mix consistency of the PNF condition is limited by the discrete output stage of its generative models. 
\begin{table}[]
\caption{SI-SDR [dB] for IRM, PNF and the proposed method (DPSS)}\label{tab1}
\centering
\begin{tabular}{l|lll|}
\cline{2-4}
\multirow{-2}{*}{}           & \multicolumn{1}{c|}{IRM} & \multicolumn{1}{c|}{PNF \cite{jayaram2021parallel}} & \multicolumn{1}{c|}{DPSS} \\ \hline
\multicolumn{1}{|c|}{speech} & \multicolumn{1}{l|}{16.25}    & \multicolumn{1}{c|}{15.93}    &  \multicolumn{1}{|c|}{\textbf{22.43}}        \\ \hline
\multicolumn{1}{|c|}{piano}  & \multicolumn{1}{l|}{13.32}    & \multicolumn{1}{c|}{12.87}    &   \multicolumn{1}{|c|}{\textbf{19.59}}      \\ \hline
\multicolumn{1}{|c|}{mix}    & \multicolumn{1}{l|}{\textbf{64.52}}    & \multicolumn{1}{c|}{38.92}    &    \multicolumn{1}{|c|}{63.34}      \\ \hline
\end{tabular}
\vspace{-0.35cm}
\end{table}
\vspace{-0.35cm}
\subsection{Listening test}
We analyzed the performance of the separation methods by two MUSHRA \cite{bs.1534-3} tests. We created a set of test mixture signals by using random items from test sets of the respective source models.\footnote{Excerpts from the test are available here: \href{https://dpss-demo.github.io/}{https://dpss-demo.github.io/}} The listeners listened separately to separated components for each of the sources in separate sessions, and they were not exposed to the mixture signals. The conditions in the tests included: a hidden reference, a 3.5~kHz low-pass anchor (LP35), and the signals extracted by the three separation methods. The anchor signals were created from the references (not from the mixture signals). 11 expert listeners participated in both listening tests. 
\par
\begin{figure}[htb]
\centering
\includegraphics[width=\linewidth]{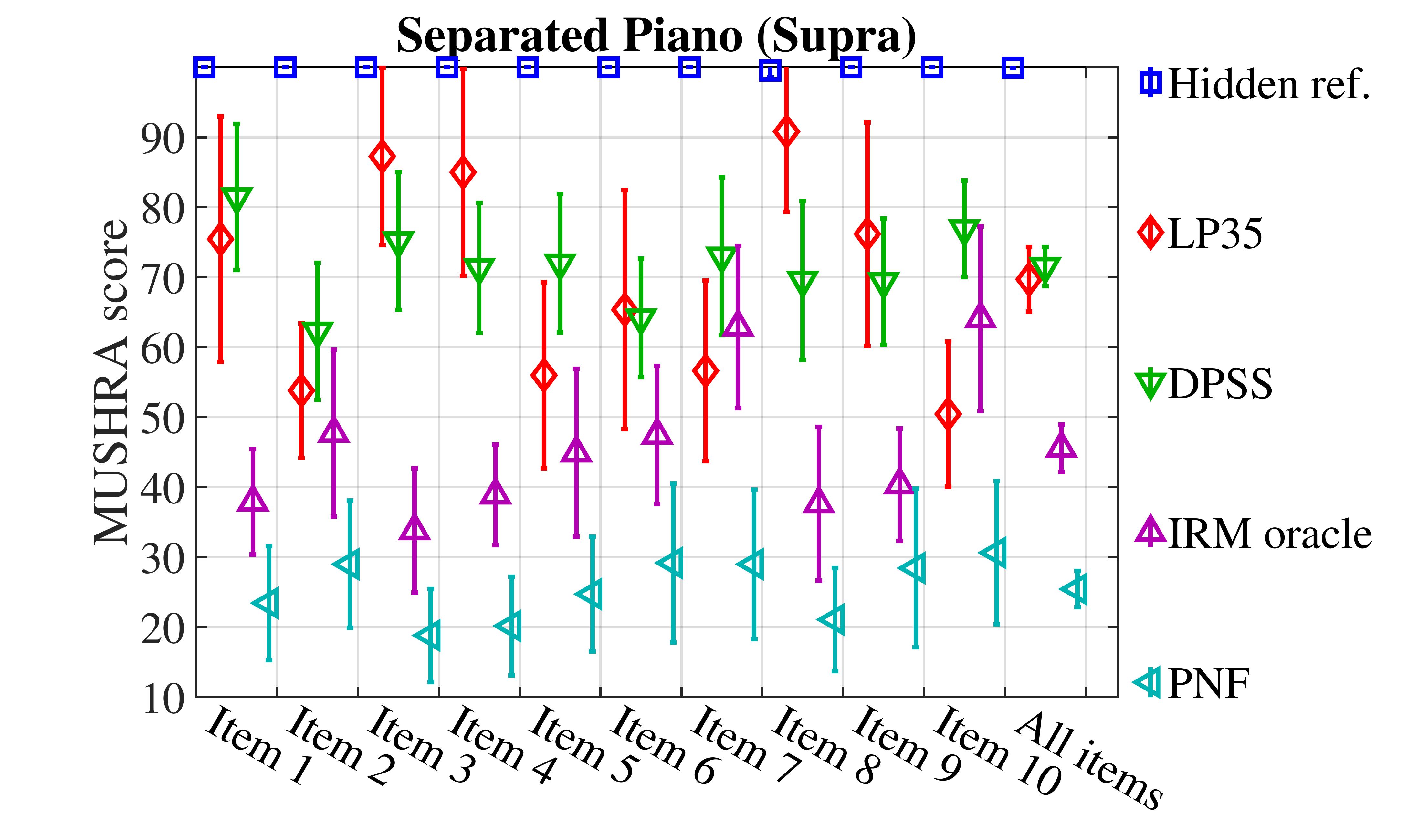}
\caption{Listening test results for Supra piano (11 listeners, 95\% confidence intervals, Student's t-distribution). \label{fig:piano}}
\end{figure}

\begin{figure}[htb]
\centering
\includegraphics[width=\linewidth]{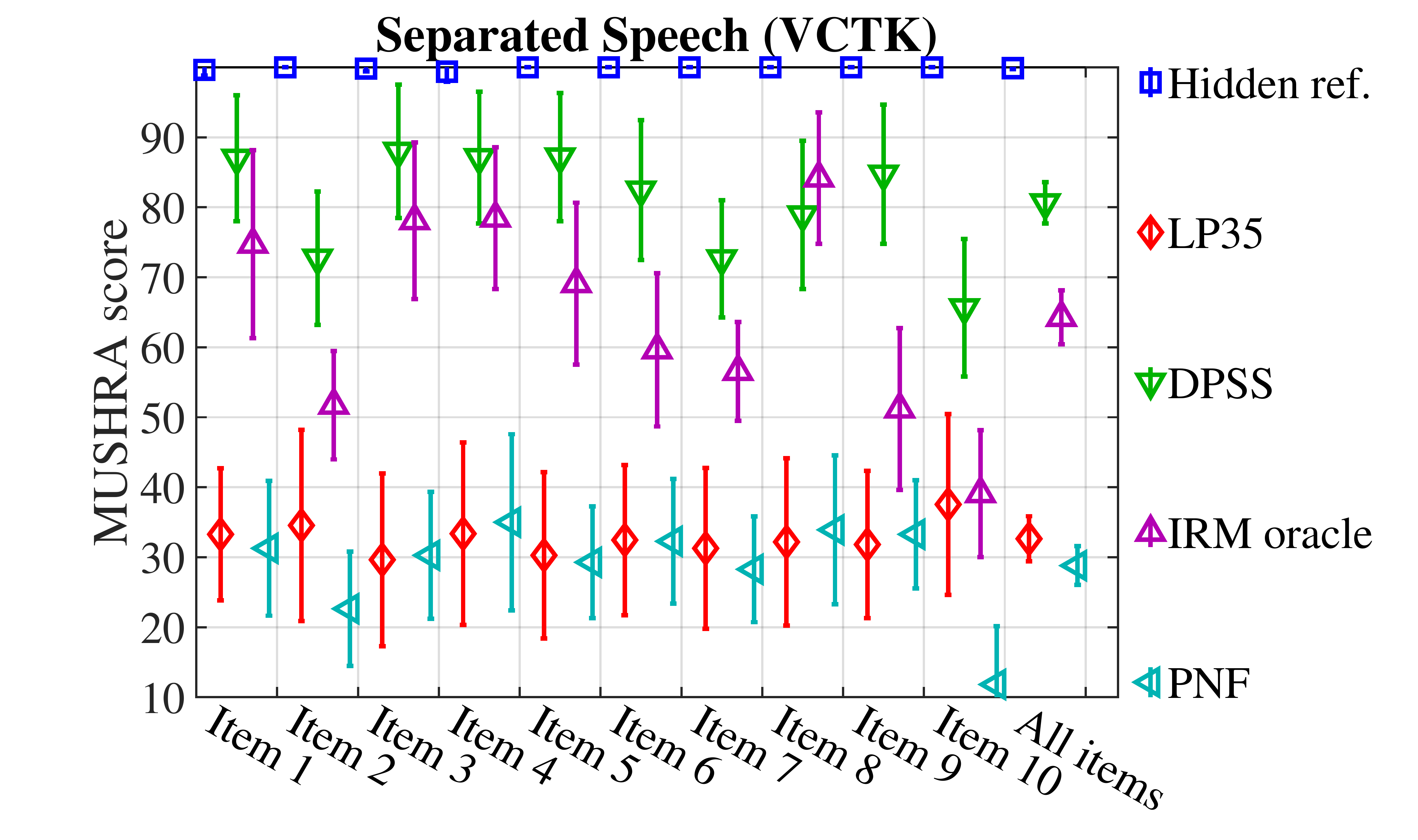}
\caption{Listening test results for VCTK speech (11 listeners, 95\% confidence intervals, Student's t-distribution). \label{fig:speech}}
\end{figure}

 The results of the listening test for separated piano are shown in Fig.~\ref{fig:piano} and the ones for separated speech are shown in Fig.~\ref{fig:speech}. For both source categories, it can be seen that DPSS provides a significant perceptual advantage, and outperforms the benchmarks. The unusually high score of the low pass anchor in the piano test is due to fact that piano signals in the Supra database have band-limited characteristics.
\vspace{-0.5cm}

\vspace{-0.5cm}
\section{Conclusion}
We conclude that preservation of source distributions can be a viable strategy for separation of audio mixtures, as it facilitates generation of plausible reconstructions of the sources. We demonstrate this advantage by performing MUSHRA tests. Furthermore, our results indicate that the proposed filterbank domain model with a continuous output stage can significantly outperform the time-domain benchmark with a discrete output stage.
 
\bibliographystyle{IEEEbib}
\begin{small}
\bibliography{refs}
\end{small}
\end{document}